\documentclass[aps,prb,superscriptaddress,showpacs,twocolumn]{revtex4-1}
\usepackage{graphicx}
\newcommand{\Co} {{\mbox{Co${}_7$(Te${}$O${}_{3}$)${}_4$Br${}_6$}}}

\begin{document}
\title{Slow magnetic dynamics and hysteresis loops of the bulk ferromagnet \Co}

\author{M. Prester} 
 \email{prester@ifs.hr}

\author{I. \v Zivkovi\'c}

\author{D. Drobac}

\author{V. \v Surija}
\affiliation{Institute of Physics, P.O.B.304, HR-10 000, Zagreb, Croatia}

\author{D. Paji\'c}
\affiliation{Department of Physics, Faculty of Science, Bijeni\v{c}ka c.32 , HR-10 000 Zagreb, Croatia} 

\author{H. Berger}
\affiliation{Institute of Physics of Complex Matter, EPFL, 1015 Lausanne, Switzerland}

\date{\today}

\begin{abstract}
Magnetic dynamics of a bulk ferromagnet, a new single crystalline compound \Co, was studied by ac susceptibility and the related techniques. Very large Arrhenius activation energy of 17.2 meV (201 K) and long attempt time ($2 \cdot 10^{-4}$ s) span the full spectrum of magnetic dynamics inside a convenient frequency window, offering a rare opportunity for general studies of magnetic dynamics. Within the experimental window the ac susceptibility data build almost ideally semicircular Cole-Cole plots. Comprehensive study of experimental dynamic hysteresis loops of the compound is presented and interpreted within a simple thermal-activation-assisted spin lattice relaxation model for spin reversal. Quantitative agreement between the experimental results and the model's prediction for dynamic coercive field is achieved by assuming the central physical quantity, the Debye relaxation rate, to depend on frequency, as well as on the applied field strength and sample temperature. Cross-over between minor- to major hysteresis loops is carefully analyzed. Low-frequency limitations of the model, relying on domain wall pinning effects, are experimentally detected and appropriately discussed.

\end{abstract}

\pacs{75.60.Ch, 75.78.Fg}

\maketitle

\section{Introduction}

Magnetic hysteresis, historically the most intensively investigated nonequilibrium phenomenon, is only recently getting better understood in all necessary details and aspects. Traditionally, the subject has a more then a century long history established in research and the applications of ferromagnetic materials in bulk- and thin-film forms. In the latter systems hysteresis relies on dynamics of pinned domain walls \cite{lee}, a subject playing a pivotal role in various fundamental aspects of magnetism \cite{nat} and its applied outcomes \cite{kru}. Nowadays, the subject  expands to nanomagnetic systems \cite{sko1}, i.e., to systems of magnetic nanoparticles \cite{fre}, as well as to molecular- and single chain magnets \cite{ses}, not necessarily involving magnetic long range order at all \cite{ses}. Besides fundamental reasons profound understanding of hysteresis and the related magnetic dynamics is urged by the demands of rapidly growing applications varying from, e.g., the ultra-high density magnetic recording \cite{fre} to magnetic hyperthermia in anti-cancer medical therapy \cite{pan}. 

Let us briefly recapitulate the main physical ingredients involved in the problem of magnetic hysteresis and the underlined dynamics \cite{fus4}. A central phenomenon that any magnetic hysteresis relies on is a dissipative spin reversal. Although an atomic-scale event, spin reversal is a subject to hierarchy of interactions rendering the hysteresis problem mesoscopic in its character. In their hierarchy, interactions generate both the equilibrium and non-equilibrium conditions for reversals. Equilibrium one follows from the appropriate free energy functional (known as micromagnetic free energy \cite{sko1}) listing the relevant energies: exchange interaction, microcrystalline anisotropy, sample-geometry-dependent magnetostatic-, and the Zeeman energy. Minimization of micromagnetic functional defines then, depending on strength of the applied field, the equilibrium single- or multi-domain state.  The archetypal Stone-Wohlfarth hysteresis model \cite{mor}, e.g., follows from such an equilibrium considerations of a single-domain sample. 

Non-equilibrium aspects in spin reversal are the consequences of energy barriers separating the metastable spin states in neighboring local minima \cite{sko}: Modeling the free energy of a hysteretic system as a multi-variable function of its magnetic degrees of freedom one sweeps, in the course of hysteresis cycle, over a complicated energy landscape. The barriers themselves rely on a number of possible sources, like a domain wall pinning on impurities/imperfections, spatial variations in local anisotropies, etc. Energy barriers are the key elements of magnetic dynamics- on basis of the Arrhenius thermal activation \cite{sko,sko1} they introduce, e.g., the relaxation time effects .

Finally, spin reversal is a cooperative phenomenon- reversing spins belong collectively to the mesoscopic or macroscopic objects, domain walls, involving large number of spins. The details are material-specific, via the peculiarities of the free energy landscape.

Out of different features characterizing hysteresis the most prominent one is coercive field- a characteristic reverse field inducing massive nucleation/propagation of domain walls. The ambition of any hysteresis model is to provide a realistic prediction for the the coercive field. Usually, the models report their results formulated as the corrections to the elementary Stoner-Wohlfarth result by taking into account dynamical features and effects \cite{sko1}. In this article we present experimental study and appropriate modelings approaching the hysteresis problem the other way around- a central investigated topic is dynamical hysteresis while the hypothetical dc hysteresis is considered as a dynamical one taken in the low-frequency limit. Opened frequency window is very important in hysteresis studies \cite{moo,pop} from the number of reasons. In coercivity studies on real samples the frequency-sensitive methods help, for example, to distinguish between the contributions originating from the Arrhenius (or N\'eel-Brown) thermal activation or from the static micromagnetism \cite{sko}.  There have been a lot of efforts invested in the attempts to reconcile the predictions of these two approaches and to identify the conditions (like temperature, frequency range) favoring one over the other \cite{sko} - this work keeps on the similar track. 

In this study we focus on the low-frequency, domain wall-related, magnetic dynamics of a new oxo-halide system $\Co$ \cite{pre,bec}. Owing to rare combination of parameters determining its magneto-dynamics, conveniently matching the dynamical hysteresis' experimental window, we show that the experimental hysteresis can be quantitatively reproduced within a simple Debye-relaxation-based model for thermally activated spin reversals. For this model to be successful in the interpretation of spin dynamics, as represented by the corresponding dynamic hysteresis loops, one has to allow the relaxation time to depend on frequency and field, as well as on temperature.

The article is organized as follows. The involved experimental techniques, sample information, as well as the main motivations, are introduced in Section \ref{SE data}. In Section \ref{reldyn} we show that domain wall dynamics just below the ferromagnetic transition (of the first order type) obeys the Debye relaxational dynamics in a wide frequency range of the presented Cole-Cole plots. We also discuss the reasons for the observed exceptionally long relaxation times and elaborate why the conditions are favorable for in-depth studies of magnetic dynamics. Experimental insight into various stages of the spin-reversal-based domain wall dynamics is presented in that section. Section \ref{dynamicloops} presents a simple model of thermal-activation-assisted Debye relaxations comparing afterwards the model predictions with the detailed experimental study of dynamic hysteresis of \Co. In Section \ref{discussion} we discuss the low-frequency limitations of the model, drawing out appropriate conclusions.

\section{Sample and experimental data}
\label{SE data}

The measurements were performed on single crystals of a new, cobalt-based oxi-halide compound $\Co$ \cite{pre,bec}. Magnetism of this system is characterized by strong magnetic anisotropy extending far into the paramagnetic temperature range. The measured bulk anisotropy relies on the strong single ion anisotropy energy of variously coordinated Co$^{2+}$ ions. At $T_{N}$=34 K an incommensurate antiferromagnetic order sets in. Its propagation vector is temperature dependent and, at $T_{c}$=27.2 K, an abrupt spin reorientation transition introduces a ferromagnetic component along $b$-axis. Standard phenomenology of ferromagnetism characterizes measurements along this axis only \cite{pre} -along other axes standard phenomenology of antiferromagnetic order, like spin-flip (or flop) transition, persists below $T_{c}$. 

Ferromagnetic component of magnetic order of $\Co$ sets in accompanied by extremely sharp and sizable peak \cite{pre,bec} in ac susceptibility. Surprisingly, the peak in imaginary susceptibility ($\chi''$) was found, depending on the strength and frequency of the applied field, significantly bigger than the peak in real susceptibility ($\chi'$). The latter result is at variance with the usual finding $\chi'' \ll \chi'$ dominating by far magnetic transitions studies of various magnetic systems in general. In an attempt to figure out why $\Co$ deviates so much from the standard behavior we realized that it represents in fact a rather unique system: its intrinsic magnetic parameters discussed below enable the entire spectrum of magnetic dynamics to unfold within the most convenient/sensitive experimental window, i.e., in the frequency range 0.05 Hz- 1 kHz. It thus enable a direct experimental access to various stages of magnetic dynamics by ac susceptibility and the related techniques. A surprising observation mentioned above, $\chi'' \geq \chi'$, a remark of big dissipation exploding in the transition, can be then very simply interpreted.

The ac susceptibility/dynamic hysteresis studies were performed by the use of a non-SQUID CryoBIND ac susceptibility system, revealing high (2 nanoemu) sensitivity, within the frequency range 50 mHz-1 kHz. The samples used were oriented single crystals cut into the rod-like geometry (typical size 0.1x1x4 mm$^{3}$), thus minimizing the value of the demagnetizing factor. The longest sample axis corresponds to the ferromagnetic $b$-axis and all measurements were performed with dc or ac field applied parallel to this axis. 

\section{Relaxational Dynamics of Domain Walls}
\label{reldyn}

\begin{figure}
\includegraphics[width=0.45\textwidth]{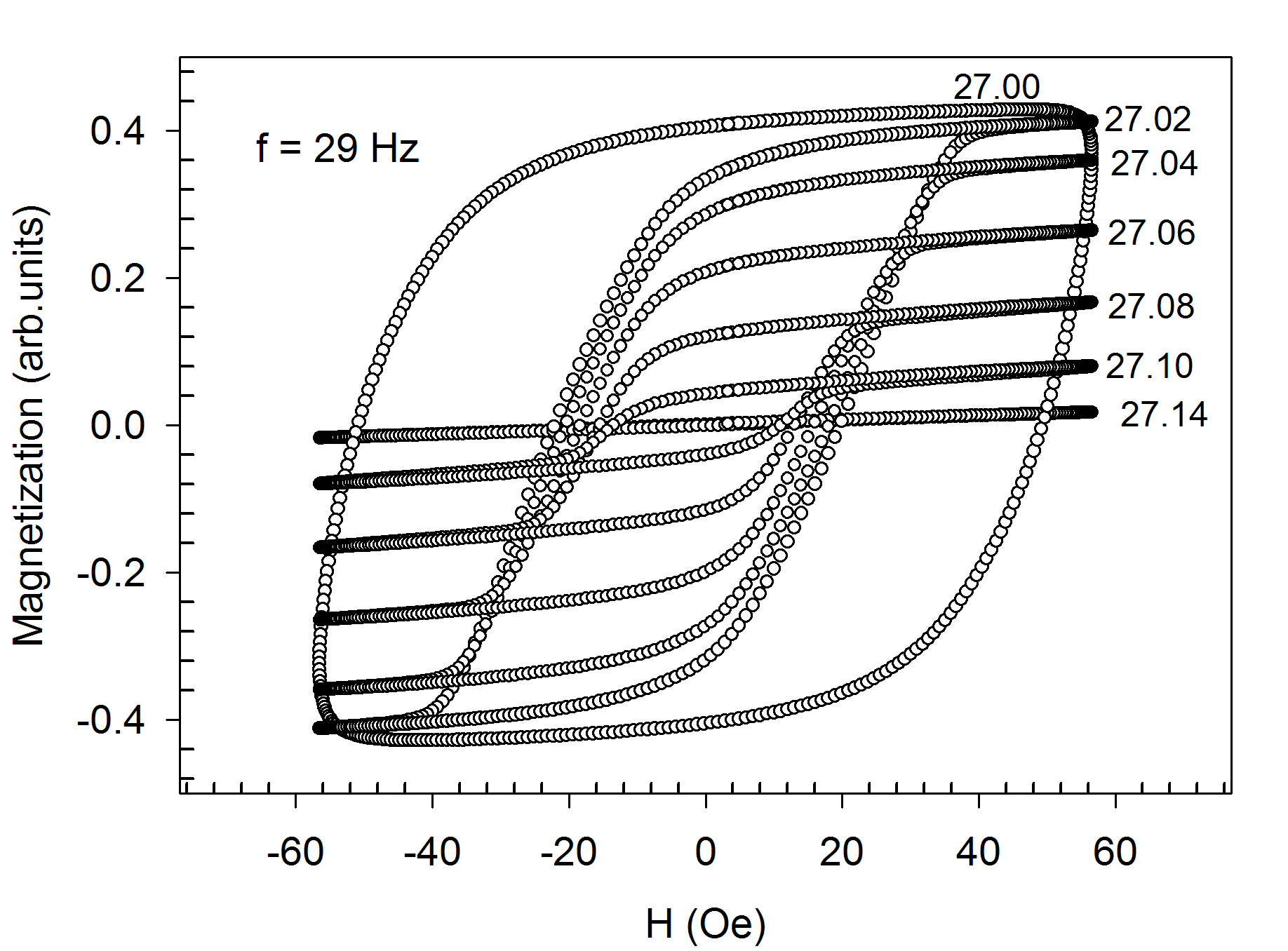}
\caption{Dynamic hysteresis loops of $\Co$ taken at several temperatures within the ferromagnetic transition. Numbers adjacent to the loops designate the temperatures. A small hystersis assymetry -a shift to the left- is commented in Section ~\ref{dynamicloops}. }
\label{firstorder}
\end{figure}

In order to clear out nature of the ferromagnetic transition at $T_c$ we first show in Fig.~\ref{firstorder} a set of dynamic hysteresis loops of $\Co$ (discussed at length in the next Sections). In the transition region these loops were taken within a narrow, 0.14 K-wide, temperature interval. Ferromagnetic order parameter -saturation magnetization- builds up its low-temperature (4.2 K) value practically within 0.1 K, illustrating the first-order character of the ferromagnetic transition \cite{kie}. This means that the magnetocrystalline anisotropy, which inherits its temperature dependence from the temperature dependent saturation magnetization \cite{zen}, is fully developed already at 27 K and barely changes by lowering temperature below 27 K. Simultaneously with anisotropy and bulk magnetization it is clear from the first principles \cite{mor} that magnetic domains and the domain walls set in and self-organize just below $T_c$, in order to keep the magnetic free energy minimal \cite{mor}. The domain pattern then responds to any changes in the parameter values by dissipative domain wall rearrangements.

%
%
\begin{figure}
\includegraphics[width=0.5\textwidth]{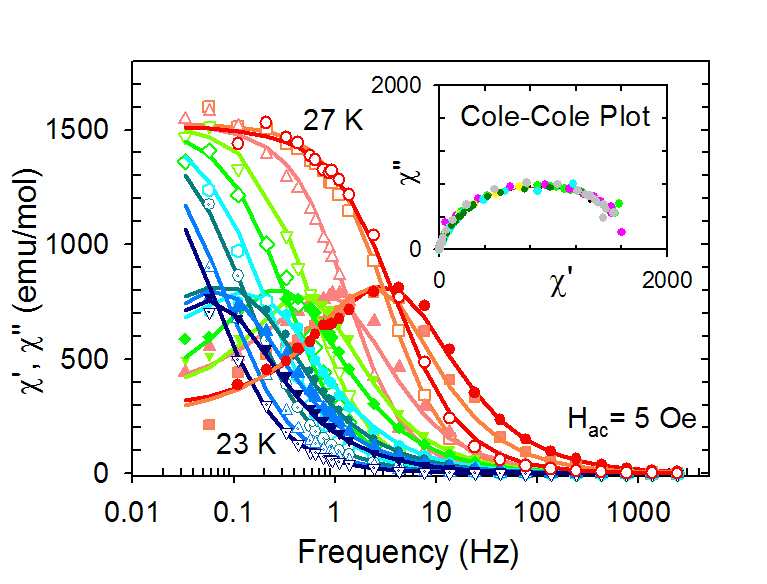}
\caption{(Color online) Frequency dependence of $\chi'$ (open symbols) and $\chi''$ (solid symbols) of $\Co$ just below $T_{c}=27.2$ K (main panel). Solid lines are fits to the standard expressions for the relaxing complex susceptibility \cite{mor}(Eq. (3)) with the addition of a prefactor in the formula for $\chi''$ to account for its asymmetry. Dynamic hysteresis and dissipation scans were measured in the tail area characterized by $\chi'' > \chi'$. Inset shows the same data from the main panel but plotted in the $\chi',\chi''$ diagram- the data are now collapsed on the single Cole-Cole semicircle.}
\label{Cole}
\end{figure}

\subsection{Cole-Cole plots}
\label{colecole}

Frequency dependence of ac susceptibility, $\chi(\omega)$, provides the most elementary insight into magnetic dynamics of ordered systems . Fig.~\ref{Cole} shows the frequency dependence of the susceptibility components $\chi'(\omega),\chi''(\omega)$ of $\Co$ crystal taken at several temperatures below $T_{c}$. This plot provides a detailed insight into magnetic dynamics of the compound. The type of response shown in Fig.~\ref{Cole} tells that the responsible magnetic subsystem, in its response to the applied field, interacts strongly with the lattice: the pattern of $\chi'(\omega,T),\chi''(\omega,T)$ curves is reminiscent of a simple Debye-type spin lattice relaxation \cite{mor}. Indeed, the $\chi'(\omega,T),\chi''(\omega,T)$ data, if presented in the form of Cole-Cole plot \cite{col}, i.e., as the functional form $\chi''(\chi')$ (Inset), all collapse on a somewhat distorted Debye semicircle (common for {\it all} temperatures within the range \cite{fus1}). Ideal semicircle, a fingerprint \cite{col} of single relaxation time $\tau$, intersects the horizontal $\chi'$-axis in the values of adiabatic $\chi_S$ (=0, $\omega \rightarrow \infty $) and isothermal $\chi_T$ ($\omega \rightarrow 0 $) susceptibility \cite{fus1}. 

\begin{figure}
\includegraphics[width=0.4\textwidth]{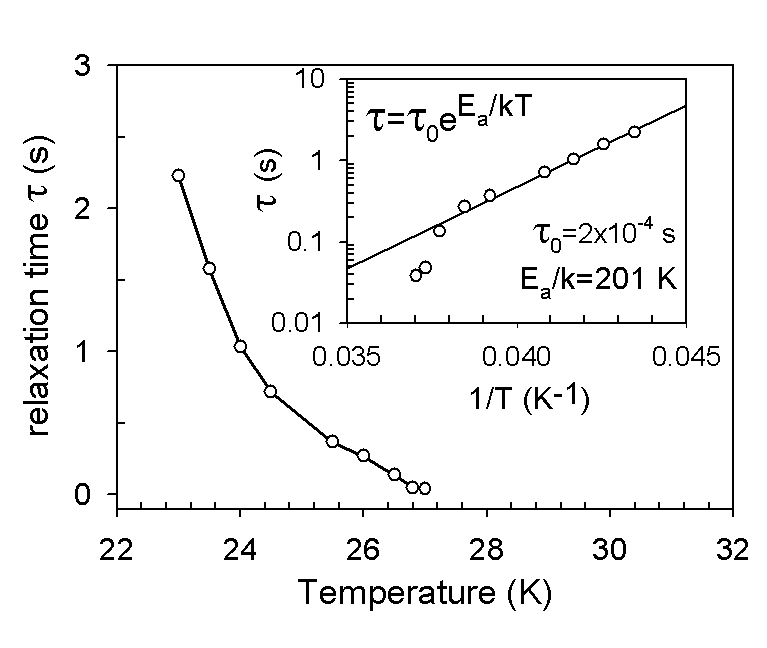}
\caption{Temperature dependence of the mean relaxation time (main panel) obeys the Arrhenius-like thermal activation with approximately temperature independent activation energy $E_a$ (inset). $E_a$ is unusually big due to strong magnetocrystalline anisotropy $K$ resulting from the strong single-ion anisotropy energy.}
\label{tau}
\end{figure}

A closer examination of $\chi'(\omega),\chi''(\omega)$ data, as well as the fact of saturation magnetization relaxing logarithmically in time (data not shown), is consistent with a distribution of relaxation times. However, in order to grasp only the basic mechanisms we hereby focus just the characteristic relaxation time $\tau$, as determined by the frequency maximizing $\chi''$. $\tau$ may be considered as the average time the magnetic subsystem, taken out of equilibrium by the application of, e.g., external magnetic field, requires to thermalize again. The temperature dependence of $\tau$ (Fig.~\ref{tau}) is obviously subject to the Arrhenius thermal activation. The involved activation energy $E_a$ represents the effective energy barrier for the thermally activated spin reversal. The reversal applies collectively to spins belonging to certain volume $V_0$ of the domain wall, an elementary reversal entity. In the context of domain wall dynamics $V_0$ is usually termed {\it Barkhausen} or {\it switching volume} \cite{sko1,moo,raq} while in superparamagnetism-related literature the related term {\it activation volume} is more common. (Although there are differences in precise physical meanings  of these terms \cite{sha} these differences are not of big importance for the main focus of this article). 

The elementary switching volume $V_0$ for $\Co$  can be readily estimated from $E_a=KV_0$, where $K$ represents the magnetocrystalline anisotropy energy. Estimating crudely $K$ from the Stoner-Wohlfarth approximation, $H_c=2K/M$, and the measured DC susceptibility values \cite{pre} for $H_c$ and $M$ one gets $V_0=5.5 \cdot 10^{-23}$ m$^3$. The edge length of the corresponding cube equals 38 nm, comprising approximately 1500 Co-ions. Activation volumes of very similar sizes have been frequently reported for, e.g., ultrathin Co-based magnetic films \cite{raq}. 

%
\begin{figure}
\includegraphics[width=0.5\textwidth]{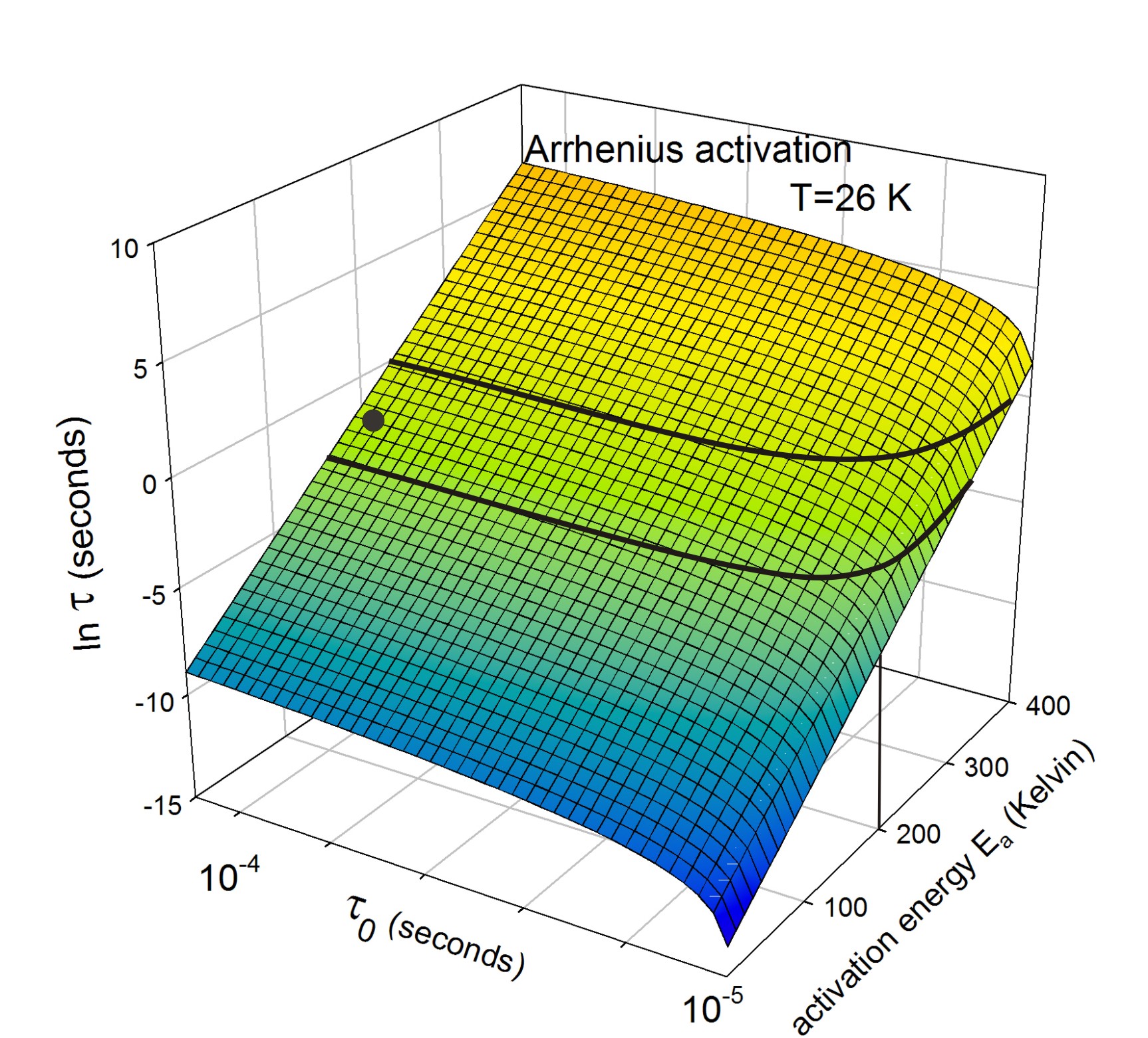}
\caption{(Color online) 
Plot of the Arrhenius thermal activation $\tau(T,E_a,\tau_0)=\tau_0 exp(E_a/k_BT)$ at fixed temperature T=26 K, in its dependence on the attempt time $\tau_0$ and the activation energy $E_a$. Thick black curves defines approximately the experimental window involved with the present ac susceptibility study. Intrinsic parameters of $\Co$ (values of $E_a$ and $\tau_0$) at T=26 K impose magnetic dynamics to take place in the position of the black dot, i.e., by coincidence, within the experimental window of readily observable area on the ln $\tau(\tau_0,E_a)$ surface. By further cooling magnetic dynamics slows down -the ln($\tau$) surface rapidly elevates- and the system's dynamics status -black dot- rapidly escapes from the experimental window area.}
\label{arrhenius}
\end{figure}

The plots shown in Fig.~\ref{Cole} witness that magnetic dynamics of $\Co$ is exceptionally well exposed to experimental studies. A conventional ac susceptibility frequency window, 50 mHz - 1 kHz, is usually too narrow to span the dynamics in its full spectrum. This is reflected by a narrow arc spanned in the $\chi''(\chi')$ plots of many investigated magnetic systems, instead of the ideal Cole-Cole semicircle \cite{cher}. In $\Co$, however, the complete spectrum of the low- to high-frequency modes gets activated within a conventional frequency window. This situation represents obviously a consequence of unusually slow relaxations characterizing magnetic dynamics of $\Co$, established already 1-2 K below $T_c$. Its slow relaxations are represented by long relaxation times $\tau(T)$ at the order of second, Fig.~\ref{tau}, determining the Arrhenius activation quantitatively. 

Physically, slow relaxations stem from the large values of both $E_a$ ($E_a/k_B=201$ K) and $\tau _0$ ($=2 \cdot 10^{-4}$ s). Let us illustrate schematically the experimental significance of these figures for studies of dynamics. Fig.~\ref{arrhenius} shows the Arrhenius surface $\tau(\tau_0,E_a)$, as it looks like on logarithmic scale for T=26 K, and the position of $\Co$, as determined by its intrinsic values of $\tau_0,E_a$. If these values were much smaller, as they usually are for large number of ferromagnets studied thus far, the position of the system at the 26 K-$\tau(\tau_0,E_a)$ surface would be deep in the blue/dark zone, i.e., well away from the experimentally achievable frequency window. With actual values of $\tau_0,E_a$ for $\Co$ its dynamic status stays inside the experimental window for temperatures in a pretty broad range, 22 K - 27 K, enabling detailed studies of domain wall dynamics. 

We relate a sizable $E_a$ primarily to the fact that ferromagnetism arises in this compound out of already macroscopically anisotropic, AF ordered, backbone \cite{pre}. In this pre-established order a strong single-ion anisotropy  plays important role, as well as in the microcrystalline anisotropy energy $E_a$ setting-in abruptly below $T_c$. As for the attempt time $\tau _0$ its experimental value is orders of  magnitude bigger than the microscopic spin-flop attempt time at the order of $10^{-9}$ s. Big $\tau _0$ is naturally attributed to collective dynamics of domain wall spins, known to take place on a slower time scale \cite{koc,kra}.

\subsection{Dissipation scans}

%
\begin{figure}
\includegraphics[width=0.37\textwidth]{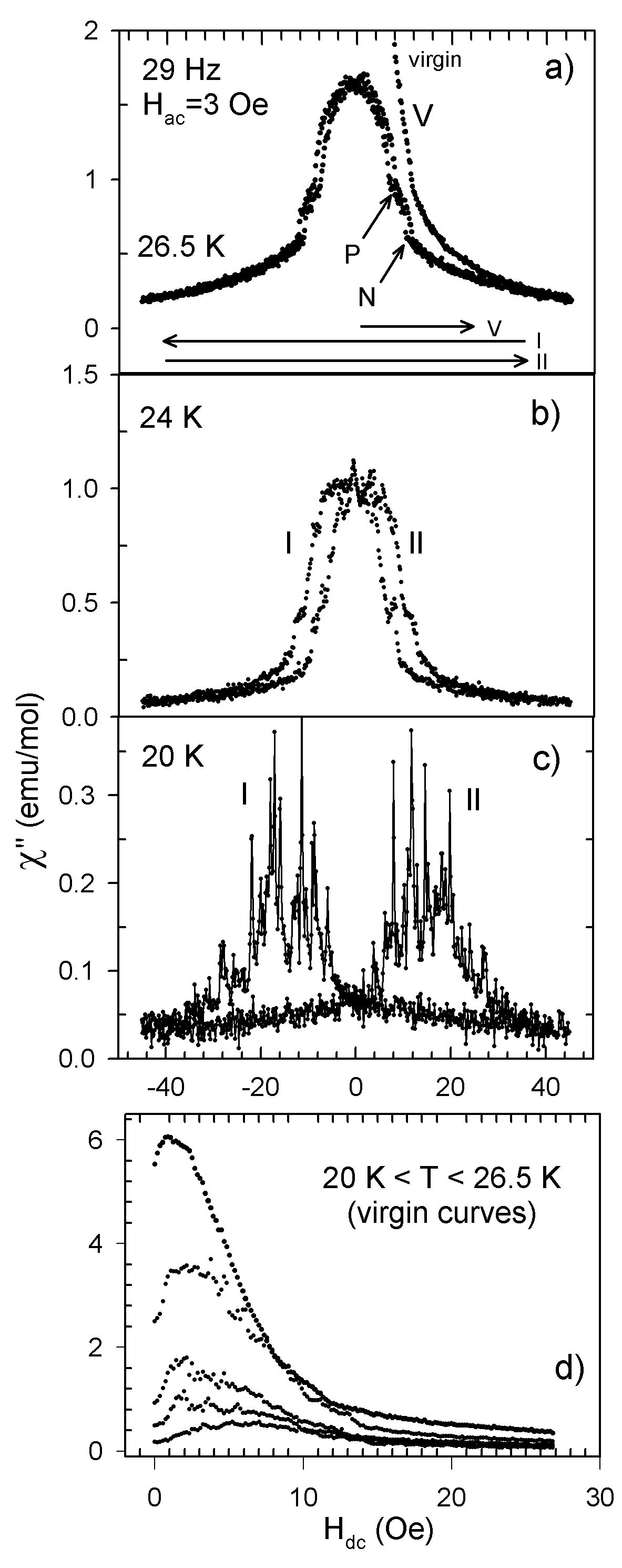}
\caption{Dissipation scans (see text), shown at different temperatures, measure the rate of spin reversals as the ramping dc field progressively ejects, and then nucleates back, the domain walls. Field sequence is indicated by horizontal arrows in (a). The points ascribed to domain walls' nucleation (N) and propagation (P) are indicated by the inclined arrows. A smooth background dissipation is attributed to incomplete spin reversal. The virgin curve dissipations are shown in (d).}
\label{butt}
\end{figure}
With the relaxation time conveniently positioned within the experimental ac susceptibility window there is an additional freedom in choosing a particular applied field's working frequency in the relaxation diagram, Fig.~\ref{Cole}. For examples, one can choose frequency on the right-hand side of the peak in $\chi''(f)$ (consistent with $\chi''\geq \chi'$, i.e., with high dissipation) and do the temperature or the dc-field scans. With such choice of frequency the temperature scan reveals explosive peak in $\chi''(T)$ at $T_c$ mentioned in the Introduction Section, which can now be reproduced by quantitative precision within a simple model based on the expressions for $\chi',\chi''$ in the Debye approximation \cite{mor} (data not shown). More interesting are the `dissipation scans' taken with the same choice of applied frequency. Dissipation scans hereby refer to recording of $\chi''$, a measure of dissipative magnetic dynamics, by the use of a small ac field amplitude superimposed over the dc field slowly ramping up and down, Fig.~\ref{butt}. DC field sweeping is accompanied by continuously changing pattern of magnetic domains through processes of domain nucleation and/or domain wall propagation. On atomic scale these processes are underlined by dissipative spin reversals. Thus the dissipative scans tell directly about the average rate of spin reversals in successive stages of domain wall dynamics, otherwise taking place in standard dc hysteresis cycling.

In the scan-starting condition of a zero-field-cooled sample at $H_{dc}=0$, necessarily composed of equal number of $+M$ and $-M$ domains, the measured dissipation was already at a high initial level, Fig.~\ref{butt}a). To achieve the highest possible rate of spin reversals application of a small temperature-dependent dc field is obviously required (Fig.~\ref{butt}d)). We speculate that a maximum in dissipation is achieved after the field supplies enough magnetic energy to overcome a weak domain wall pinning energy. In the virgin curve a rapid dissipation decay indicates ejection of the domain walls out of the sample rendering it single-domain and revealing just residual dissipation afterwards. In the field back-swing reversed domains start nucleating again and further dissipation evolves through propagation of the domain walls as well. Both processes are integrated into the `dissipation dome' structure, Fig.~\ref{butt}. The temperature-dependent positions of the maxima can naturally be identified with the coercive field $H_c(T)$. The $H_c(T)$ values, being small immediately below $T_c$ but well-defined, are in good quantitative agreement with the dc-SQUID $H_c(T)$ data \cite{pre}. 

The structure of the dissipation dome tells not only on the activation of the dissipative nucleation/propagation modes but also on the importance of thermal activation. After being expelled from the sample the domain re-nucleation relies on overcoming the anisotropy-related energy barriers for spin reversal of the activation volume. This obviously takes place with the most frequent rate at $H_c$ revealing, naturally, the field strengths being negative in sign (in the field first back-swing). But one also notes, for loops taken at T=26.5 K, 24 K, that the very onset of nucleation takes place already for positive field values. Domain nucleation in positive field can only rely on thermally-induced reversals of the activation volumes. The latter finding provides an independent argument favoring the importance of thermal activation in studies of magnetic dynamics in $\Co$. Fig.~\ref{butt} also shows that, unlike the domain walls ejection from the virgin state, the $H_c$-build up is not a smooth process. Obviously, at lower temperatures it progresses via erratic, perhaps avalanche-based, mechanism which is not well understood as yet (and is not elaborated in this work any further).

\section{Dynamic hysteresis loops of $\Co$}
\label{dynamicloops}

Now we present a more detailed study of the evolution of magnetic dynamics in $\Co$. One way to do it, as shown in the studies on ultrathin films \cite{che,kle}, would consist of collecting a set of the Cole-Cole plots for different $H_{ac}$.  In our work, however, the magnetic dynamics was studied more directly, by measuring the induction-type dynamic hysteresis responding to the sinusoidal applied field $H_{ac}=H_{0}\:$sin$\: \omega t$. The $M-H$ loops were determined at different temperatures and studied as functions of the field amplitude $H_{0}$ and frequency $\omega$. 

\begin{figure}
\includegraphics[width=0.48\textwidth]{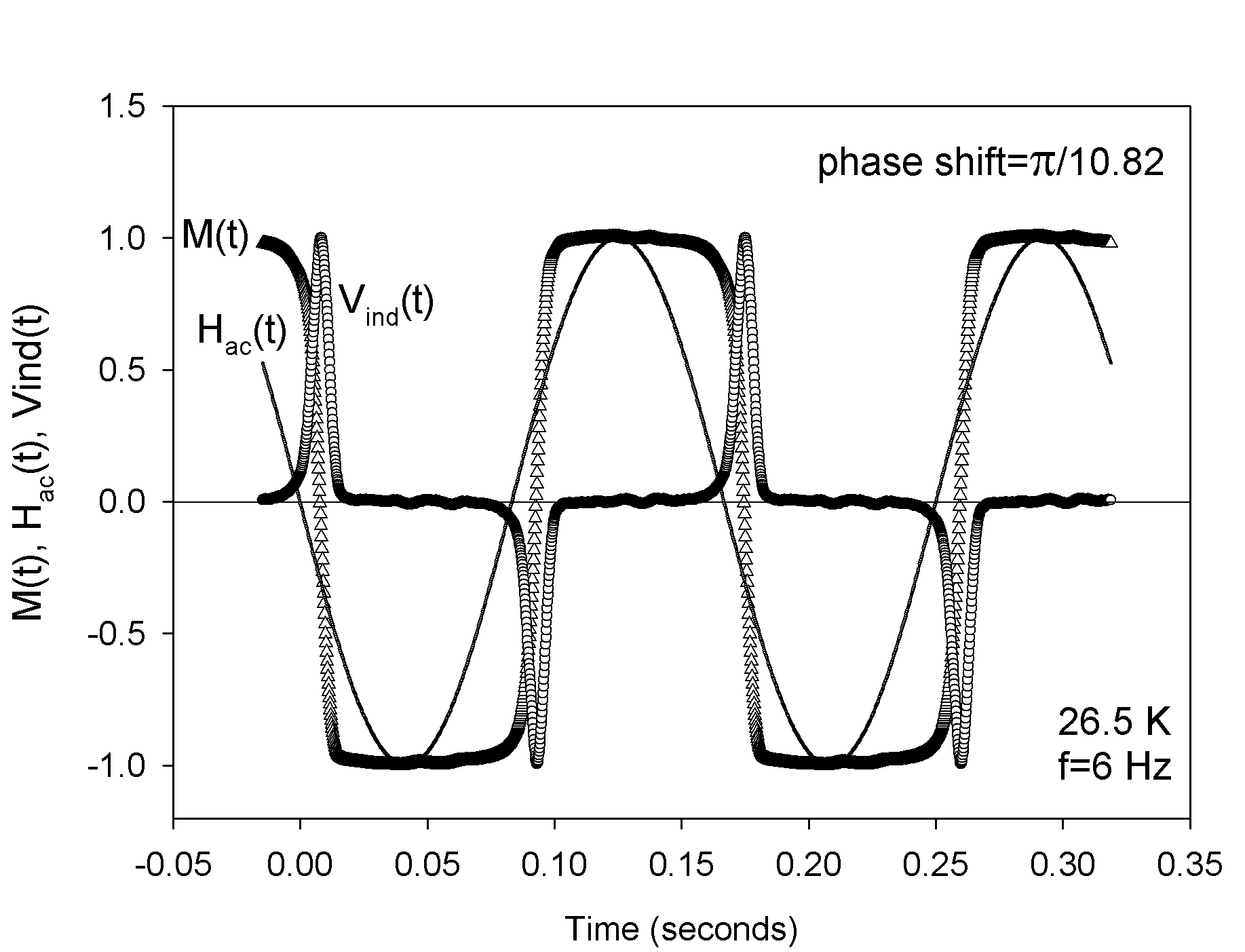}
\caption{An example of dynamic hysteresis measurement on $\Co$ in the time domain. Magnetic field pulse $H_{ac}(t)$ and the induced voltage  $V_{ind}(t)$ (circles) are measured and recorded directly while magnetization $(M(t)$ (triangles) is obtained by numerical integration of $V_{ind}(t)$. The phase shift between $H_{ac}(t)$ and $M(t)$ brings a direct information on the involved magnetic dynamics (see, text). All quantities are plotted scaled by their maximal values.}
\label{timedomain}
\end{figure}

Magnetization $M(t)$ of the sample was determined by fast digital acquisition of voltage $V_{ind}(t)$, induced in the highly balanced secondary coil of our ac susceptometer by the application of sinusoidal current through the primary. By virtue of Faraday law, $V_{ind}(t) \sim d\phi/dt \sim dM/dt$. Fig.~\ref{timedomain} shows the relevant signals in the time domain. 

The period of the applied pulse was varied within the range 1 ms-1 s enabling studies of magnetic dynamics on the 1 Hz-1 kHz frequency scale. In order to improve statistics the identical subsequent strings of $V_{ind}(t)$ data were appropriately averaged. Depending on the field frequency averaging takes 1-100 s, enabling continuous measurements of appropriately averaged $V_{ind}(t)$ data in the course of a slow temperature drift. The field signal $H_{ac}(t)$ was taken by recording voltage on the standard resistor connected in series with the primary coil. In numerical elaboration of the data, taken afterwards, magnetization $M(=\int V_{ind}(t)dt)$ was obtained by numerical integration and subtraction of the background. $M(t)$ is then plotted versus $H_{ac}(t)$ to get the hysteresis loops in their temperature evolution. 

\begin{figure}
\includegraphics[width=0.48\textwidth]{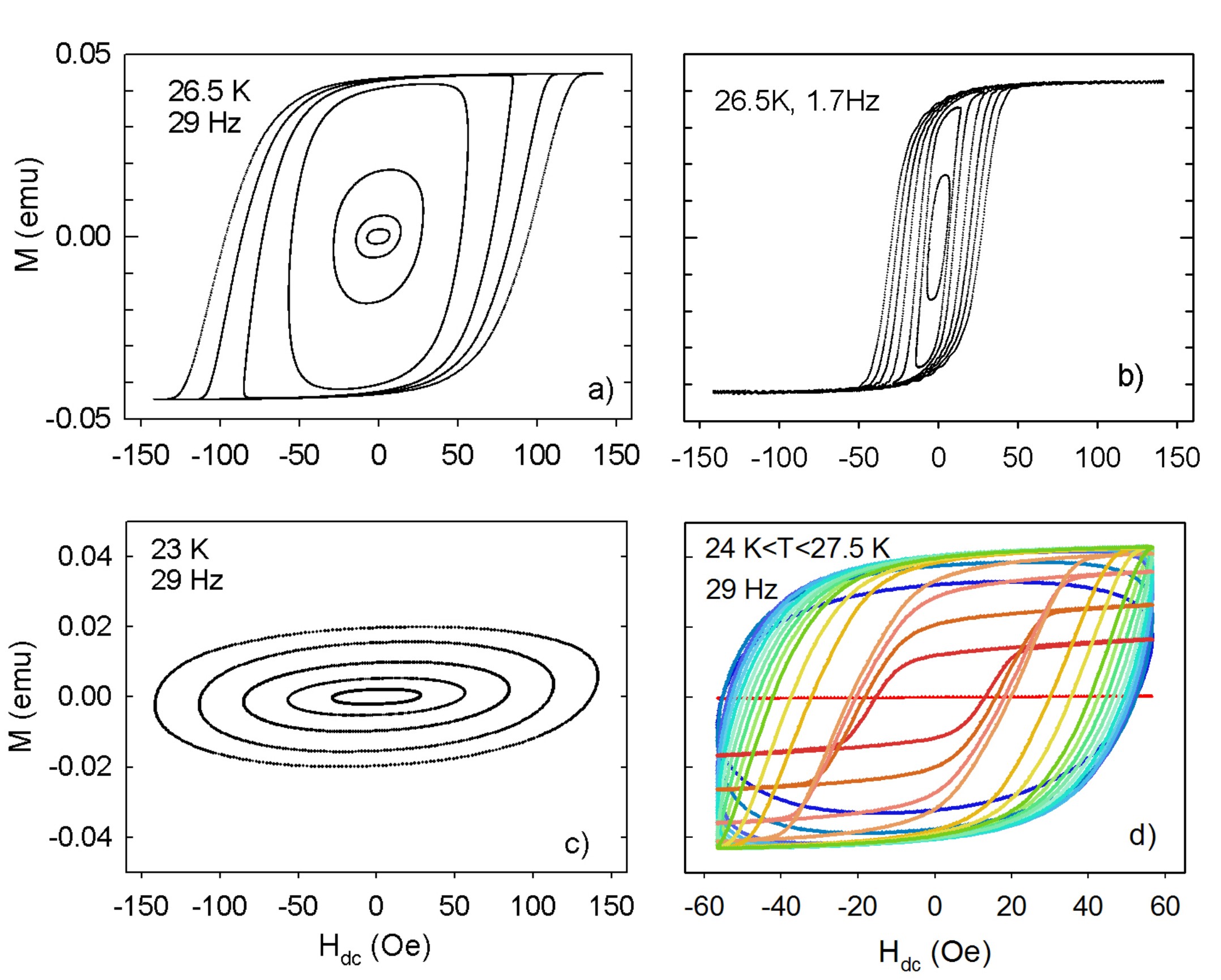}
\caption{(Color online) Experimental, induction-type dynamic hysteresis on a 2.9 mg single crystal taken with sinusoidal magnetic field pulses varying in amplitudes (5-100 Oe RMS). Calibration to absolute units was performed by the use of dc susceptibility results. The pulse widths were $t=$34 ms and $t=$0.59 s, shown in panels (a),(c),(d) and (b), respectively. At higher frequencies (29 Hz) the loops -generic ellipses- are characteristically horizontally aligned (panel c)) reflecting fulfillment of the condition $\chi''\gg \chi'$ (see, text). At lower temperatures the domain wall propagation does not reach the sample boundaries within the available field amplitudes. All aspects of magnetic dynamics are revealed in the set of 29 Hz hysteresis loops at different temperatures, shown in (d), including the collapse of order parameter for $T$ passing through $T_c$. The loops are also characterized by small asymmetry commented in the text.}
\label{hys}
\end{figure}

In Fig.~\ref{hys} a set of representative dynamic hysteresis data on $\Co$ is shown. At fixed chosen frequencies and measurement temperatures a sequence of loops were taken in increasing magnetic field amplitudes, defining a broad range of conditions for dissipative spin reversals.

Referring to the loop dependence on $H_{0}$ there is obviously a crossover between the two generic loop types, i.e., between the minor loops for small values of $H_{0}$ and, in increasing $H_{0}$, the major loops revealing characteristic `whiskers'. As it is generally well understood \cite{nat} minor loops correspond to domain wall dynamics taking place in the limit of infinite sample size while the major ones correspond to saturation to the sample-size-limited total magnetization. In the whisker-like hysteresis regions domain walls are entirely ejected out of the sample thus any dynamics dimes out in the whisker region. During the cycle dynamics reappears again within the reversed magnetic field swing. It is then underlined by back-nucleation and propagation of domain walls. The activation energies involved with minor and major loops are not identical; while the domain wall propagation is mostly involved with the minor loops the major loops relies, besides propagation, on the domain wall nucleation \cite{lab,raq}. Thus the dynamical features of minor and major loops are expected to be at least quantitatively different. The details of these differences, showing up in a continuous transformation of minor- into major loops in increasing $H_{0}$, are presented and analyzed in this study.

Hereby, we pay particular attention on the phenomenology of minor loops, attracting recently a pronounced renewed interest \cite{pop,kob}. As illustrated in Fig.~\ref{hys} (c) and elaborated below the minor loop is a generic ellipsis, the parameters of which depend sensitively on the applied field magnitude $H_{0}$, its frequency $f$, and the sample temperature $T$. To the best of our knowledge the detailed elaboration of the latter aspects of magnetic dynamics has not been presented so far for any ferromagnetic system. For this purpose we first introduce a simple model for minor loops. Then, in the next section, we show the model to be in sound qualitative and quantitative agreement with the presented experimental investigations of magnetic dynamics of $\Co$.

\subsection{Model for minor hysteresis loops}

For the sake of interpretation of the experimental loops we first invoke the simplest phenomenological concept of hysteretic behavior in general: macroscopic `response' of a hysteretic system (i.e., magnetization in our case) lags behind the `external probe' (i.e., instantaneous value of the applied magnetic field) for the phase shift $\phi$. Thus the magnetization $M(t)$, lagging behind the field $H(t)$, reads
\begin{eqnarray} 
\label{eq1}
H(t)=H_{0} \:{\rm sin} \:\omega t, 
\nonumber\\ 
M(t) = M_0 \:{\rm sin} (\omega t - \phi) = M_0 \:{\rm cos} \phi \:{\rm sin} \:\omega t -
\nonumber\\ 
-M_0 \:{\rm sin} \phi \:{\rm cos} \:\omega t.
\end{eqnarray}
If one then defines the in-phase and the out-of-phase susceptibility components as $\chi'=(M_0/H_0)\:$cos$\:\phi$ and $\chi''=(M_0/H_0)\:$sin$\:\phi$, respectively, the magnetization becomes 
\begin{equation} 
\label{eq2}
M(t)=\chi' H_0 \:{\rm sin} \:\omega t - \chi'' H_0 \:{\rm cos}\:\omega t. 
\end{equation}
The phase shift is now determined as the ratio of the two susceptibility components, tan$\phi=\chi''/\chi'$.

\begin{figure}
\includegraphics[width=0.45\textwidth]{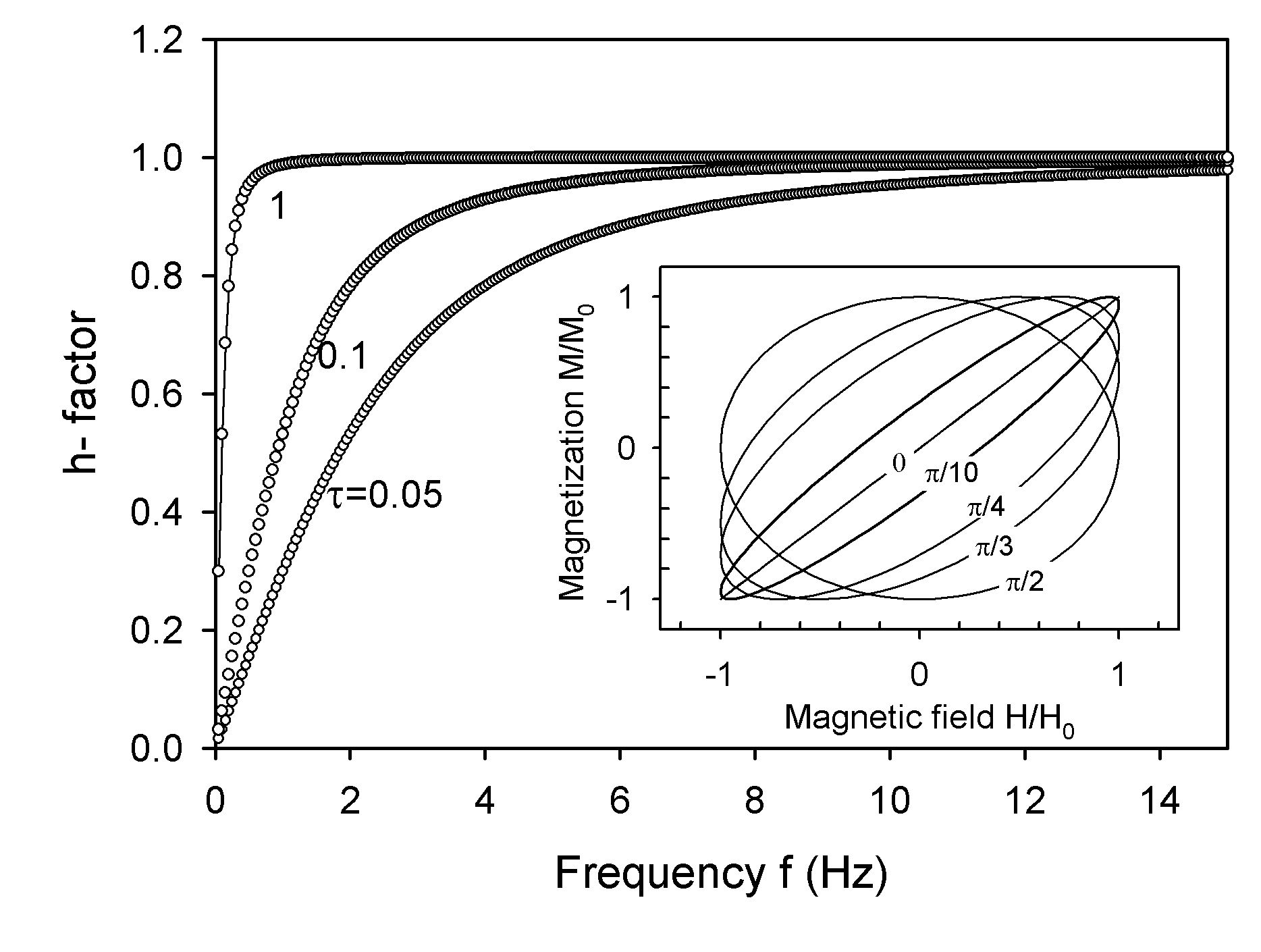} 
\caption{Coercive field factor $h(\omega,\tau)\equiv \omega \tau / \sqrt{1+ \omega^2 \tau^2}$ for three representative values of $\tau$, effective in thermal activation of $\Co$ at approximately 24 K, 26.5 K and 27 K, have been selected for the plots. Inset: Magnetization vs. applied field from Eq.(\ref{eq1}), normalized to respective amplitude values $H_0,M_0$, for several chosen values of the phase shift parameter $\phi$ (numbers adjacent to each loop).}
\label{hcvsomega}
\end{figure}

At this stage it is instructive to visualize several characteristic $M(H)$ plots for several chosen values of $\phi$. In Inset to Fig.~\ref{hcvsomega} we show the corresponding $M(t),H(t)$ values (for $t$ treated as an implicit variable in Eq.(\ref{eq1})), revealing the hysteretic $M-H$ loops. The loops are represented by generic ellipses tilted with respect to the $H$-axis for an angle determined by the phase factor $\phi$. 

The latter simple model for hysteresis loops is very general as no assumption for the involved type of magnetization (irrespective if it is of paramagnetic, ferromagnetic, superparamagnetic,..., origin), nor for the involved phase-lagging mechanism, has been introduced. It however strictly applies only to the case of linear magnetic response, i.e., to $M_0$ being independent on $H_0$, a condition ascribed to minor loop regime.  We focus our elaboration now to the specific dynamics of domain walls in $\Co$. As shown in Section~\ref{reldyn} the latter dynamics fits nicely the model of Arrhenius-Brown-N\'eel thermal activation. Within the latter model the dissipative spin reversal is thermally activated thus it is naturally to assume that thermal activation plays important role in hysteresis formation as well; hereby, we think of dc-hysteresis as the dc-limits of the frequency-dependent dynamic hysteresis. The general model of dynamic hysteresis becomes specific once the phase parameter $\phi$ gets defined. We therefore introduce the phase relationship inherent to the domain-wall-related relaxational dynamics. In its simplest Debye form (single relaxation time $\tau$) and within the Casimir-du Pre approximation this dynamics is represented by the well-known non-resonant magnetic susceptibility form \cite{mor} :
\begin{eqnarray} 
\label{eq3}
\chi'=\chi_S+ \frac{\chi_T - \chi_S}{1+ \omega^{2} \tau^{2}}
\nonumber\\ 
\chi''=\frac { (\chi_T - \chi_S) \omega \tau}{1+ \omega^{2} \tau^{2}}.
\end{eqnarray}
In this simple linear response theory approach all physics of the system is condensed in the phenomenological constant $\tau$, which, in turn, can depend on a number of microscopic and macroscopic parameters. Remaining consistently at the phenomenological level one can put now the phase shift $\phi$, a quantity decisive for the shape of general hysteresis (Inset to Fig.~\ref{hcvsomega}), on physical grounds by relating it to the relaxation time $\tau$: Taking for adiabatic susceptibility $\chi_S=0$ (in accordance with our experimental data) one gets \cite{fus} from Eq.3 that the Debye-Casimir-du Pres relaxation implies tan$\phi=\chi''/\chi'=\omega \tau$, thus $\phi=\arctan \omega \tau$. 

One naturally asks to what extent the experimental details of dynamical hysteresis of $\Co$ shown in Fig.~\ref{hys} can be described by such a model. As a demanding test one may look for coercive field and its dependence on the strength and the frequency of the applied magnetic fields. In this study, the term `coercive field' is used strictly in the context of dynamical hysteresis loops- the corresponding values differ a lot from the `technical' coercive fields, usually extracted from the dc magnetic hysteresis on samples reaching magnetic saturation. Similarly as for the dc hysteresis we define the (dynamic) coercive field $H_c$ simply as $M(H_c)=0$. As one reads from Eq.(\ref{eq1}) magnetization extinguishes whenever $\omega t = \phi$, i.e., disregarding the periodicity, at instant $t_c= \phi / \omega $. The corresponding field value (i.e., coercive field) is:

\begin{eqnarray} 
\label{eq4}
H(t_c)=H_c(\omega, \tau)=H_0 \:{\rm sin} \:\omega t_c=H_0\:{\rm sin} \phi =
\nonumber\\ 
=H_0 \:{\rm sin}\:{\rm arctan \: \omega \tau}=H_0 \:{\rm sin}\:{\rm arcsin} \frac{\omega \tau}{\sqrt {1+ \omega^2\tau^2}}=
\nonumber\\ 
=H_0\frac{\omega \tau}{\sqrt {1+ \omega^2 \tau^2}}.
\end{eqnarray}

Eq. (4) represents the operative result of this simple model which can be tested now on the experimental dynamic hysteresis curves. 

Depending on the frequency of the applied field there are obviously two limits. In the limit of a long measurement period with respect to the relaxation time (hysteresis at low frequencies,  $\omega \tau \ll 1$) Eq. (4) implies $H_c\approx H_0 \omega \tau$.  Hysteresis axis is then approximately aligned along the line $M(H)=M_0 H$ in the $M-H$ plot, cf. Inset to Fig.~\ref{hcvsomega}. In the opposite limit of short measurement period (high frequencies, $\omega \tau \gg 1$) Eq. (4) implies $H_c\approx H_0$ and the phase lagging approaches its maximum value $ \phi \approx \pi/2$; hysteresis axis is then horizontally aligned, cf. Inset to Fig.~\ref{hcvsomega}. The main panel of Fig.~\ref{hcvsomega} shows the frequency dependence of the factor $h(\omega,\tau)\equiv \omega \tau / \sqrt{1+ \omega^2 \tau^2}$, which rules, at fixed $\tau $,  the frequency dependence of $H_c$ in this model (Eq.(\ref{eq4})).

Invoking now our experimental dynamic hysteresis (Fig.~\ref{hys}) one easily notes that there is at least qualitative accordance: A change in hysteresis alignment, from a steep almost vertical one (panel (b)) to the horizontally aligned one (panel (c)), can naturally be interpreted as transformation in measuring conditions introducing a cross-over between the low- to high-frequency limits in the presented model.

Summarizing the physical ingredients built-in in this dynamic hysteresis model  we note that these ingredients are identical to those underlying the standard spin-lattice relaxation model~\cite{mor} for $\chi',\chi''$, usually presented in the form of the Cole-Cole plots (see, section \ref{colecole}). Thus dynamic hysteresis studies, interpreted within this model, could represent a method alternate and/or complementary to the usual spin-lattice $\chi',\chi''$-route. For example, Eq. (4) offers, in its low-frequency limit, the relaxation rate $\tau$ to be extracted from the measurements of $H_c$. Especially in measurements looking for temperature dependence of $\tau$ dynamic hysteresis route might turn out to be advantageous.

Referring again to all our experimental dynamical hysteresis (Figs. 1,7,10) one notes a small but systematic shift of the hysteresis body (at the amount of a few-Oersted) to the left. Thus, if one defines $H_c^-$ and $H_c^+$ as the coercive fields for the field-down and the field-up hysteresis branches, respectively, one notes $\mid H_c^- \mid >\mid H_c^+\mid $. One reason for the loop asymmetry could be a b-axis component of the Earth field, not compensated in the presented experiments. Hysteresis shift of typically 3-5 Oe would however be difficult to interpret on basis of the interference with the Earth field (approximately 0.3 Oe). Another possible reason for the loop asymmetry might rely on the fact that a subject b-axis ferromagnetism in $\Co$ represents just a component of the global antiferromagnetic order. We thus speculate that a scenario similar to the one valid in the exchange-biased structures, revealing characteristic hysteresis shifts, might be effective. More work would however be necessary to clarify the observed loop asymmetry.

\subsection{Experimental dynamic hysteresis loops of $\Co$: Dependence on frequency and amplitude of measuring fields}

\begin{figure}
\includegraphics[width=0.4\textwidth]{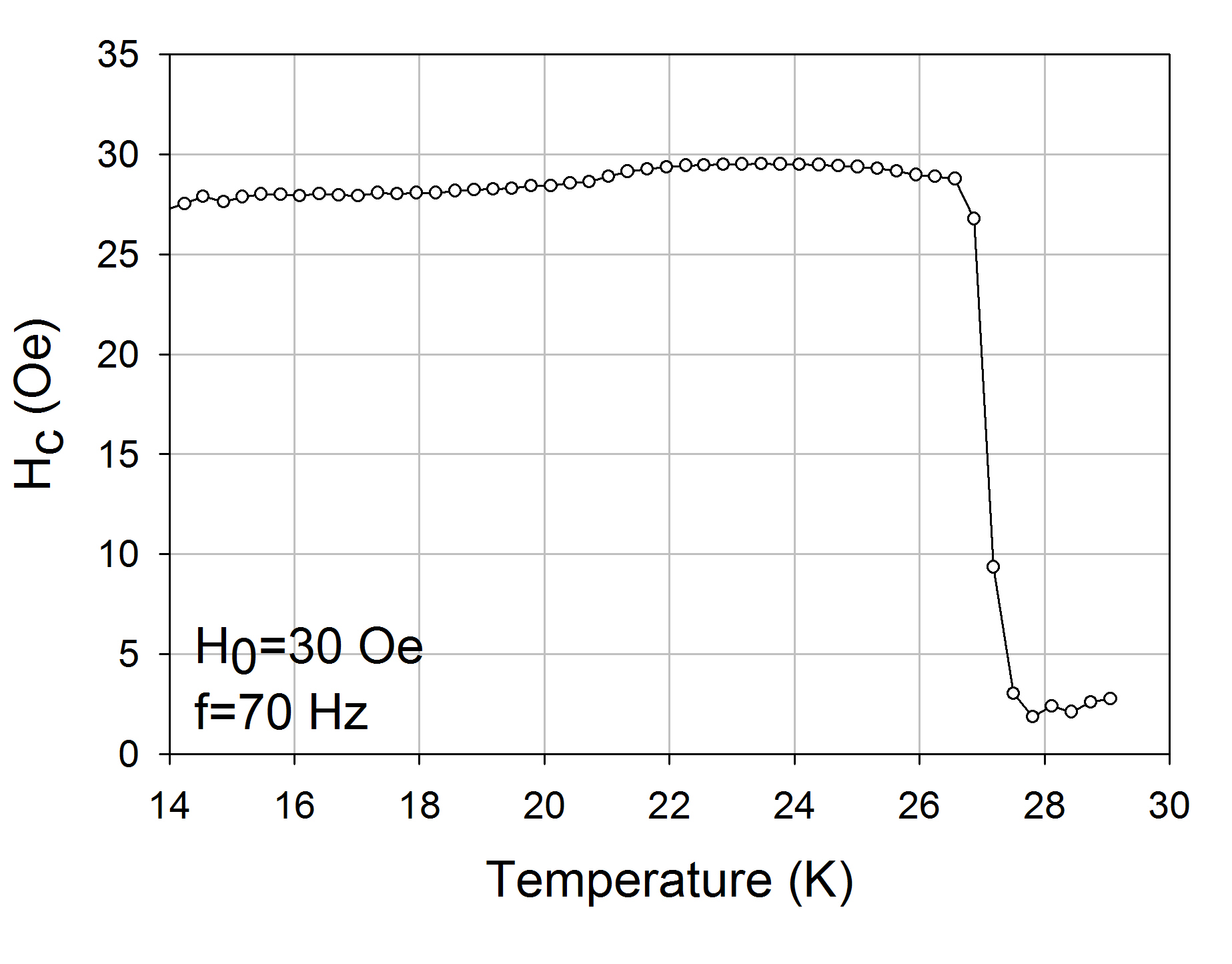} 
\caption{Temperature dependence of $H_c$ as determined from the dynamical hysteresis curves collected during a slow temperature drift (like those shown in Fig.~\ref{hys}, panel d)). Some sample dependence has been observed in the $H_c(T)$ measured on different samples. Solid line is guide to the eye.}
\label{fighcvst}
\end{figure}

In view of the model presented above, particularly its Eq. (4), we present now a more quantitative study of the dynamic hysteresis results. Eq. (4) is obviously symmetric with respect to $\tau$ and $\omega$ - this model's coercive field is a function of the product $\omega\tau$ product only. For the fixed $\omega$ one can first check the coercive field dependence on relaxation time $\tau$. In $\Co$ $\tau$ is strongly temperature dependent and, below $T_c$, obeys the Arrhenius activation law $\tau(T)=\tau_0 exp(E_a/k_BT)$, with $\tau_0={\rm 2 \times 10^{-4}}$s and $E_a/k_B=$201 K. Within a temperature window of several Kelvins the relaxation time $\tau$ grows by cooling (Fig.~\ref{tau}) from zero (at $T_c$) to several seconds. Thus for the applied field in the typical frequency range 1 Hz - 1 kHz the high frequency limit $\omega \gg 1/\tau $ will in any case be achieved soon below $T_c$. In view of Eq. (4) one expects steep increase of $H_c$ below $T_c$ followed by a saturation of $H_c$ at $H_0$, the amplitude of the applied sinusoidal field. This is exactly what has been observed for one set of measuring parameters ($f,H_0$), Fig.~\ref{fighcvst}. Saturation of $H_c$-values at the level of the field amplitude $H_0$ is accompanied by leveling of the hysteresis-ellipsis axis  with the H-axis (cf. Fig.~\ref{hys}d) ), just as in the elementary model of dynamic hysteresis in the limit $\phi \rightarrow \pi/2$ (Inset to Fig.~\ref{hcvsomega}).

\subsubsection{Frequency dependence of $H_c$}

\begin{figure}
\includegraphics[width=0.4\textwidth]{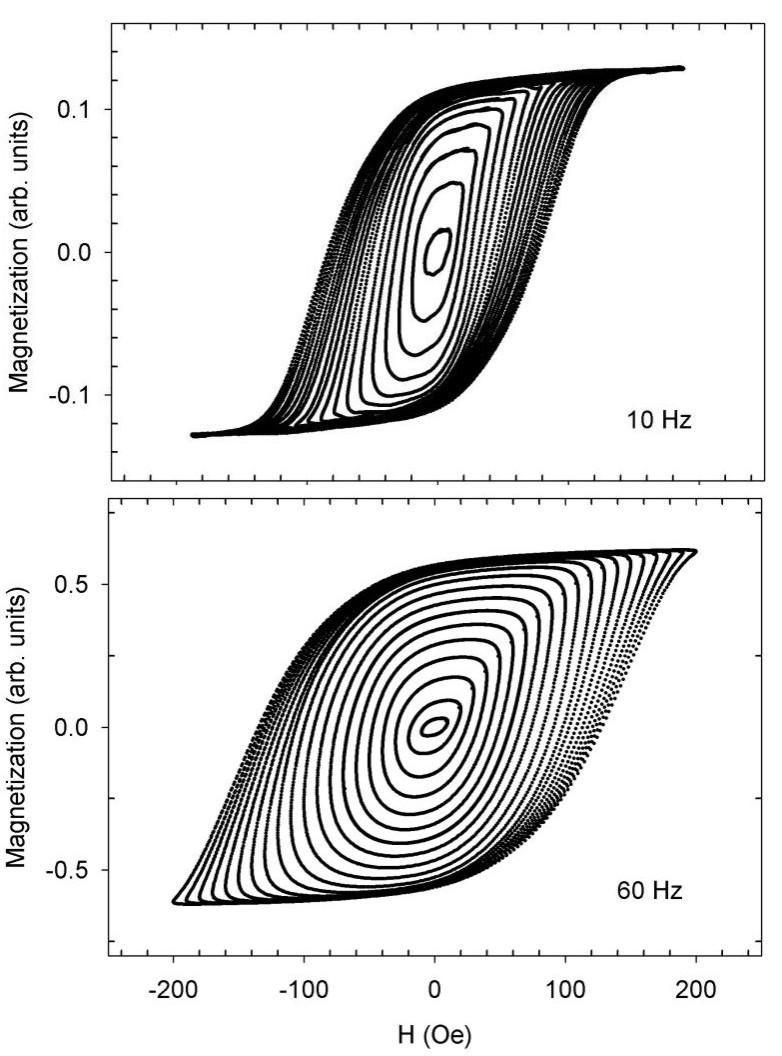} 
\caption{Dynamic hysteresis on m=2.4 mg $\Co$ sample at 26.5 K in sequence of applied field amplitudes $H_{0}$(10-200 Oe, 10 Oe steps), at the two measuring frequencies, 10 Hz, and 60 Hz. To save space the sets of hysteresis loops taken at number of other frequencies are not shown.}
\label{fign1}
\end{figure}

For the sake of a more detailed verification of the dynamical susceptibility model a large number of the measuring sequences were taken at several stabilized temperatures below $T_c$. A measurement sequence was typically taken on a zero-field-cooled sample at fixed applied field frequency, performing acquisition of hysteresis loops using gradually increasing field strengths. Before taking another sequence, at different frequency, the sample was routinely heated above $T_c$ and cooled back to the temperature of the previous sequence. Fig.\ref{fign1} shows the two representative sets measured at two frequencies in a broad range of applied fields $H_{0}$. One immediately notes that in measurements taken at 10 Hz the saturated, whisker-like, hysteresis were clearly achieved while at 60 Hz only the distorted minor loops were present. The recorded sets of dynamical hysteresis loops enable a quantitative data analysis in terms of frequency and field strength treated either as a variable or as a parameter. Here we present the coercive field dependence first on frequency and then on the field strength/magnitude, exploring the range of applicability of the model's predictions (as summarized by Eq. (4)).

\begin{figure}
\includegraphics[width=0.5\textwidth]{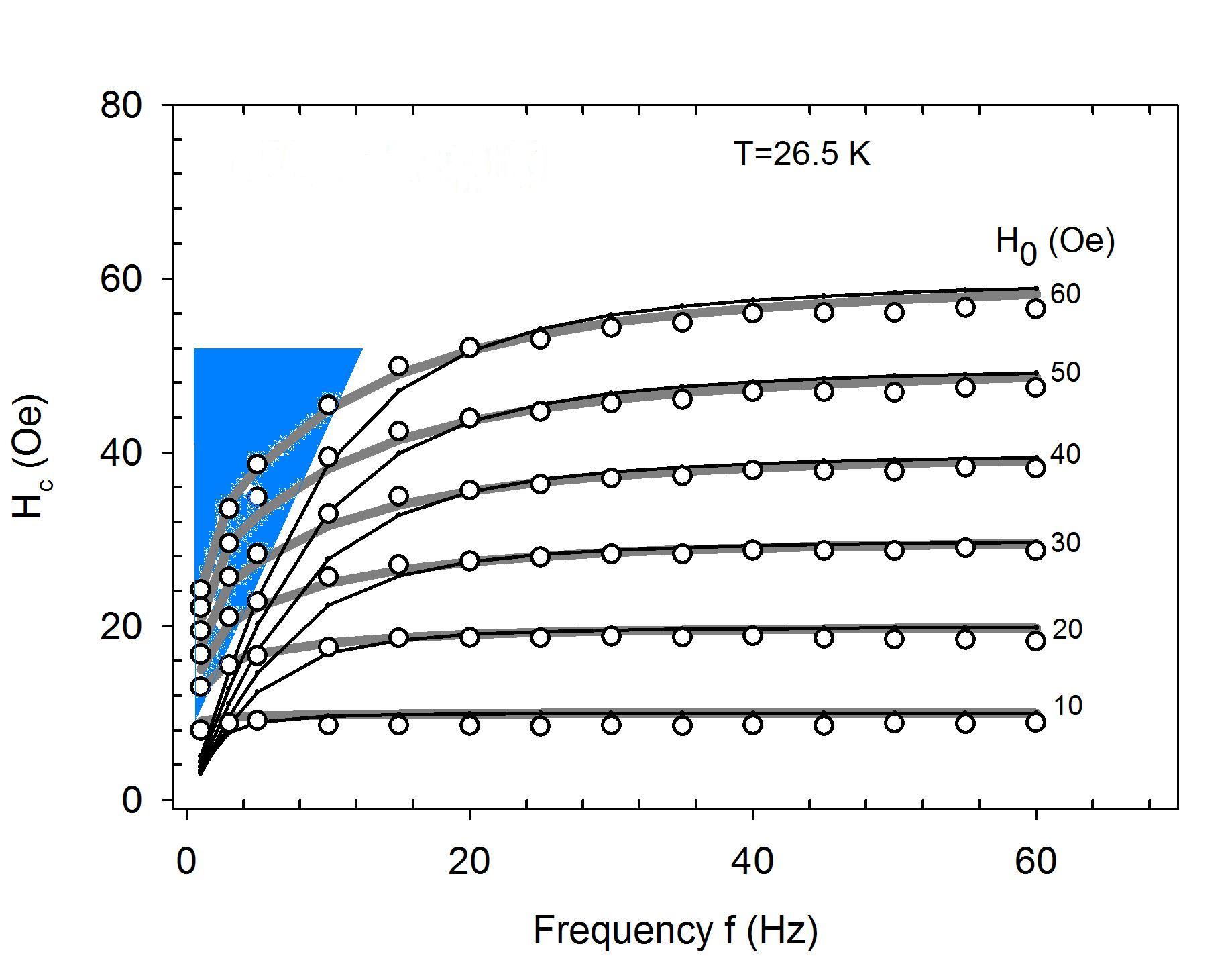} 
\caption{(Color online) Dependence of $H_c$ on frequency $f$ at fixed $H_0$. The experimental $H_c$ points (circles) are derived from measurements partially shown in Fig.\ref{fign1} (for two frequencies only). Thin solid lines represent the predictions of the model, Eq.\ref{eq4}, treating $\tau$ as a fit variable depending on $H_0$ only. Thick solid lines represent fits to Eq.\ref{eq4} treating $\tau$ as a function of $H_0$ \textit{and} $f$  in the form of a simple product, $\tau(f,H_0) \equiv \tau_f(f)\tau_h(H_0)$. Domain wall pinning takes important role in the triangle area.}
\label{fign3}
\end{figure}

\begin{figure}
\includegraphics[width=0.4\textwidth]{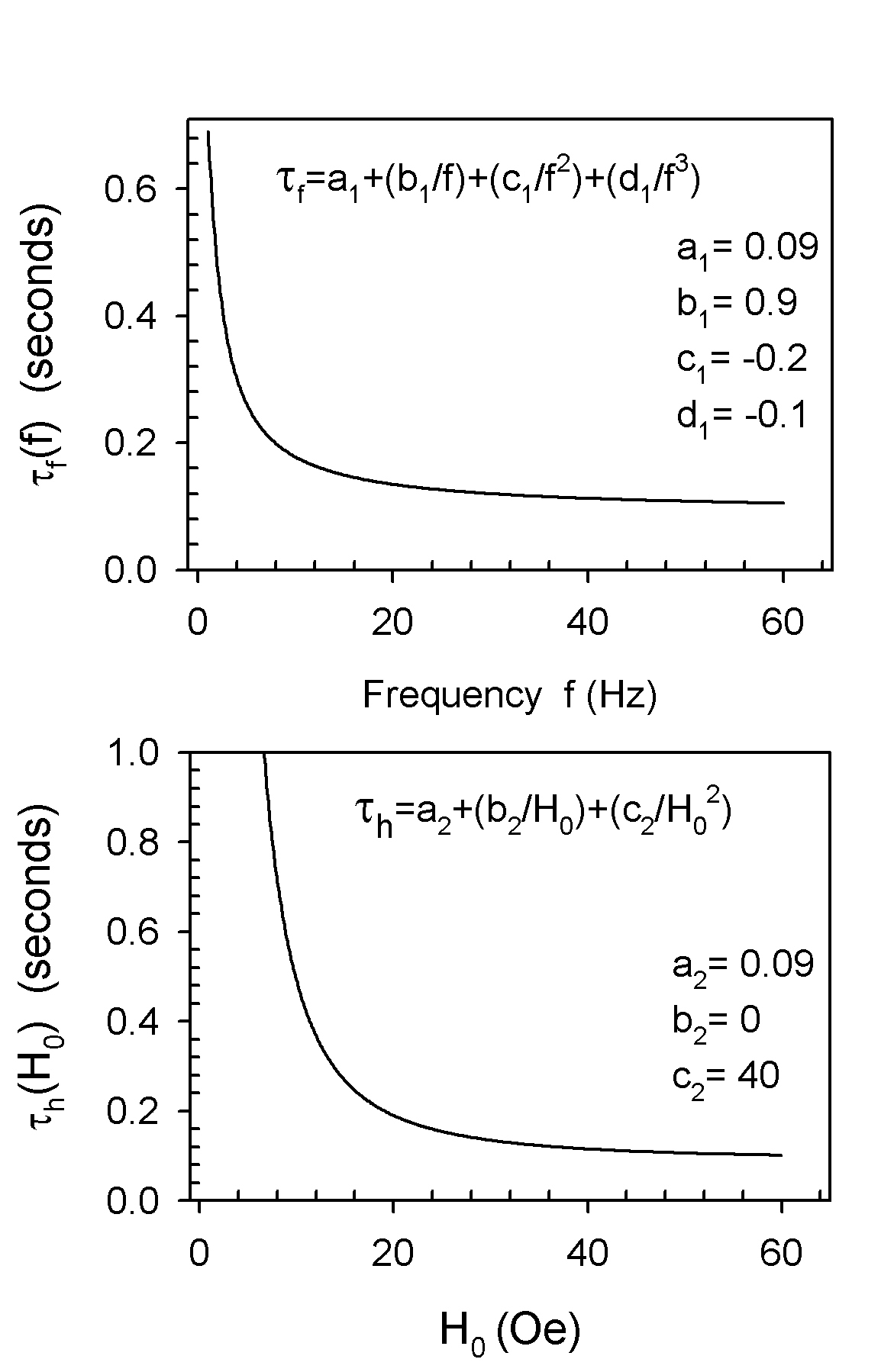} 
\caption{Plots of the single-variable functions $\tau_f(f)$ and $\tau_h(H_0)$ used in heuristic expression  for $\tau$, $\tau(f,H_0)=\tau_f(f) \tau_h(H_0)$. These functions are used to interpret the experimental $H_c(f)$  and $H_c(H_0)$ points (Fig.\ref{fign3}) and Fig.\ref{fign2}, respectively)}.
\label{fign4}
\end{figure}

Fig.\ref{fign3} shows experimental dependence $H_c$ on frequency, to be compared with the model's predictions, illustrated in Fig.\ref{hcvsomega}. By increasing frequency up to 60 Hz and above, the expected result, $H_c=H_0$, achieved at all
fields in the range, is clearly consistent with the high-frequency limit of the model. However, agreement within the full frequency range (solid lines in Fig.\ref{fign3}) was achieved only after the relaxation rate $\tau$ was assumed to represent not only a function of temperature but also a function of frequency and magnitude of the applied field, thus $\tau=\tau(T,f,H_0)$. The physical background behind this assumption is discussed in the next section. The fitting procedure was performed in two steps. First, the best possible agreement between the experimental $H_c(f)$ points and Eq. (4) was found assuming $\tau$ to depend only on $H_0$ in the form of a simple second-degree polynomial in $H_0$ (thin solid lines in Fig.\ref{fign3}). While for frequencies above 15-20 Hz the thin lines, representing the $H_c(f)$ values calculated from Eq. (4), are in sound agreement with the experimental $H_c(f)$ points there are large deviations in the low-frequency range. Frequency dependence of $\tau$ was therefore introduced as well, by assuming a heuristic expression for $\tau$, $\tau(f,H_0)=\tau_h(H_0)\tau_f(f)$. With $\tau_f(f)$ and $\tau_h(H_0)$ both in the form of second-degree polynomial there is obviously a full agreement between the experimental points and the prediction of the model for the $H_c(f)$ dependence extending in the whole frequency range. The analytical forms for $\tau_h(H_0)$ and $\tau_f(f)$, found to give the best fit, are shown in Fig.\ref{fign4}. We note that the $\tau_h(H_0)$ depends to a large extent on the applied $H_0$ range- if a wider range were applied the fit parameters of $\tau_h$ would be very different. 

\subsubsection{Dependence of $H_c$ on the field magnitude: minor- to major- hysteresis loop crossover}

\begin{figure}
\includegraphics[width=0.48\textwidth]{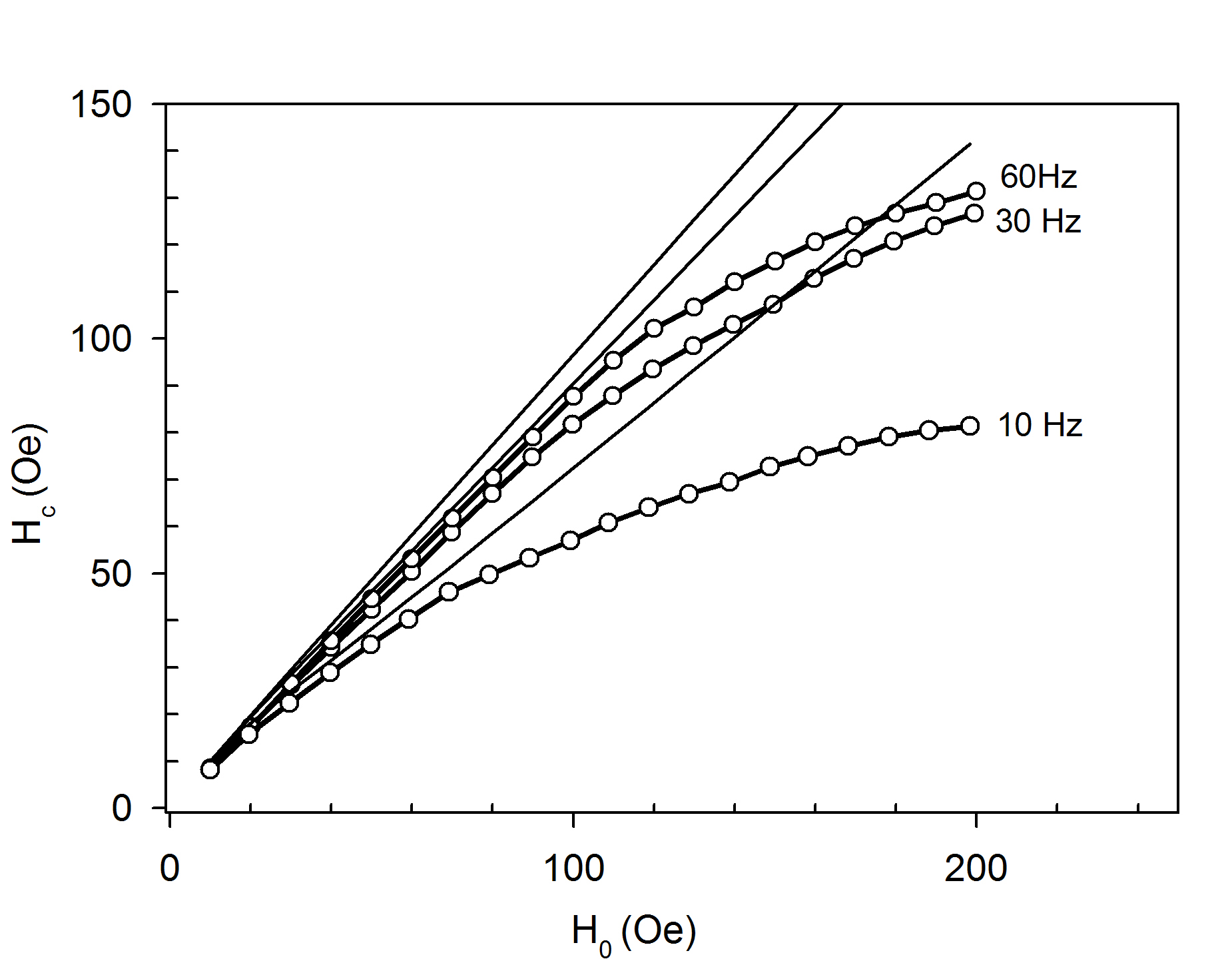} 
\caption{Coercive field $H_{c}$ as a function of applied field amplitudes $H_{0}$ for three measuring frequencies. The plots are derived from measurements shown in Fig.\ref{fign1}. Solid lines represent the $H_c$ values calculated from Eq.\ref{eq4} using the same $\tau(f,H_0)$ functional dependence involved in the thick-line fit specified in Fig.\ref{fign3}, with the parameters listed in Fig.\ref{fign4}.}
\label{fign2}
\end{figure}

Focusing primarily the minor loop mechanisms we were focusing, up to now, the low-field regime (far from magnetic saturation). Now we want to extend our results to higher fields. Instead of presenting the frequency dependence of $H_c$ it is instructive to plot the dependence of $H_{c}$ on the applied field magnitude $H_{0}$ for fixed frequencies. Fig.\ref{fign2} shows the  $H_{c}(H_0)$ dependence in the broad range (up to 200 Oe) of $H_{0}$ at the three frequencies, as extracted from Fig.\ref{fign1} data. Within this field range transformation of minor- into the major-hysteresis loops inevitably takes place (Figs. \ref{hys},\ref{fign1}). Deviation of the experimental points in Fig.\ref{fign2} from the present minor loop model (solid lines) can naturally be interpreted as a consequence of the latter transformation.
At the highest frequency (60 Hz) the functional form $H_c=H_c(H_0)$ is not only linear but the simple equality $H_{c}=H_{0}$ holds, up to high values of $H_{0}$. Such a behavior is fully consistent with the model's prediction in the high-frequency ($\omega \tau \gg 1$) limit (horizontally aligned loop). For measurements at lower frequencies, however, $H_{c}(H_0)$ deviates from linearity and tends to saturation already in much smaller values of $H_{0}$. By comparison with master plots Fig.\ref{fign1} it is clear that the $H_c(H_0)$ saturation tendency reflects the stages of the minor-to-major loop transformation. Whatever the frequency, as long as $H_0$ value grants the whisker-free (i.e., the minor loop) regime  $H_c(H_0)$ keeps its linear slope; for measurements at 60 Hz it stays true up to 100 Oe or more (cf. Fig.\ref{fign2}). At lower frequencies, however, deviation from the linear $H_c(H_0)$ slope sets-in; lower the frequency stronger the deviation from linearity. Referring to master plots Fig.\ref{fign1} one notes that the quasi-saturation level in $H_c(H_0)$ dependence, diminishing with frequency, is correlated with the constantly growing size of the dynamics-free whisker region in increasing $H_{0}$.

\section{Discussion and conclusion}
\label{discussion}

Result that the experimental curves become consistent with the present model of thermal-activation-enhanced Debye relaxations, simply by letting the relaxation time $\tau$ to depend on field and frequency (Fig.\ref{fign4}), deserves some additional comments and considerations. It is particularly interesting to discuss the range of applicability of this model in the frequency domain by investigating the matching of the two coercive fields, dynamic (subject of this work) and the static one.

Let us first summarize some general knowledge on the domain wall dynamics underlying hysteresis loops. During a hysteresis field cycle the collective spin reversal is subject to different dynamical regimes ruled by different microscopic mechanisms. As a result, within a single cycle the domain walls are subject either to wall propagation or the domain nucleation \cite{raq}. Dynamics of these modes involves different relaxation times \cite{raq,moo} and their activation depends not only on the actual magnetic field value (and its history) but also on frequency of the applied field (field ramping rate). It has been suggested \cite{moo}, for example, that the mechanism of wall propagation (with its specific relaxation time) dominate over the mechanism of domain nucleation at low frequencies (and vice versa at high frequency), introducing a minimum in the frequency dependence of coercive field at the characteristic -cross-over- frequency \cite{moo}.

Let us apply this understanding to analysis of our results/model. Within a single field cycle of our sinusoidal field there is, as discussed above, processes involving very different `instantaneous' relaxation times. These relaxation times, accumulated during a cycle, directly influences properties of our experimental dynamic hysteresis (like $H_c$) through fundamental relationship for the phase shift in our model, $\phi =\arctan \omega \tau$. We can discuss now the observed field/frequency dependence of our effective $\tau$ in more details. The central dissipative process, collective spin reversal, can be modeled as the collective spin jumps over the energy barrier separating the two neighboring energy minima. In micromagnetic models \cite{sko} most of magnetic dynamics' features (like the relaxation times) evolve from properties of the corresponding energy landscape. The barrier heights, thus the whole landscape, depend sensitively on the applied magnetic field. The role of the applied ac field is therefore twofold: First, it continuously redistributes population of the `up' and `down' domains within the field cycle to keep track with the Boltzmann distribution function. Redistribution takes place through the dissipative movement of domain walls. Second, magnetic field reshapes, through local reduction of the barrier heights, the energy landscape inside which the collective spin reversal takes place. Field-induced reshaping of the energy landscape has a direct consequence on the effective relaxation times \cite{sko} - shallower the barriers, shorter the relaxation time. Thus even weak fields involved with minor loops reduces the effective $\tau$, in agreement with our results for $\tau_h(H_0)$. 
\begin{figure}
\includegraphics[width=0.48\textwidth]{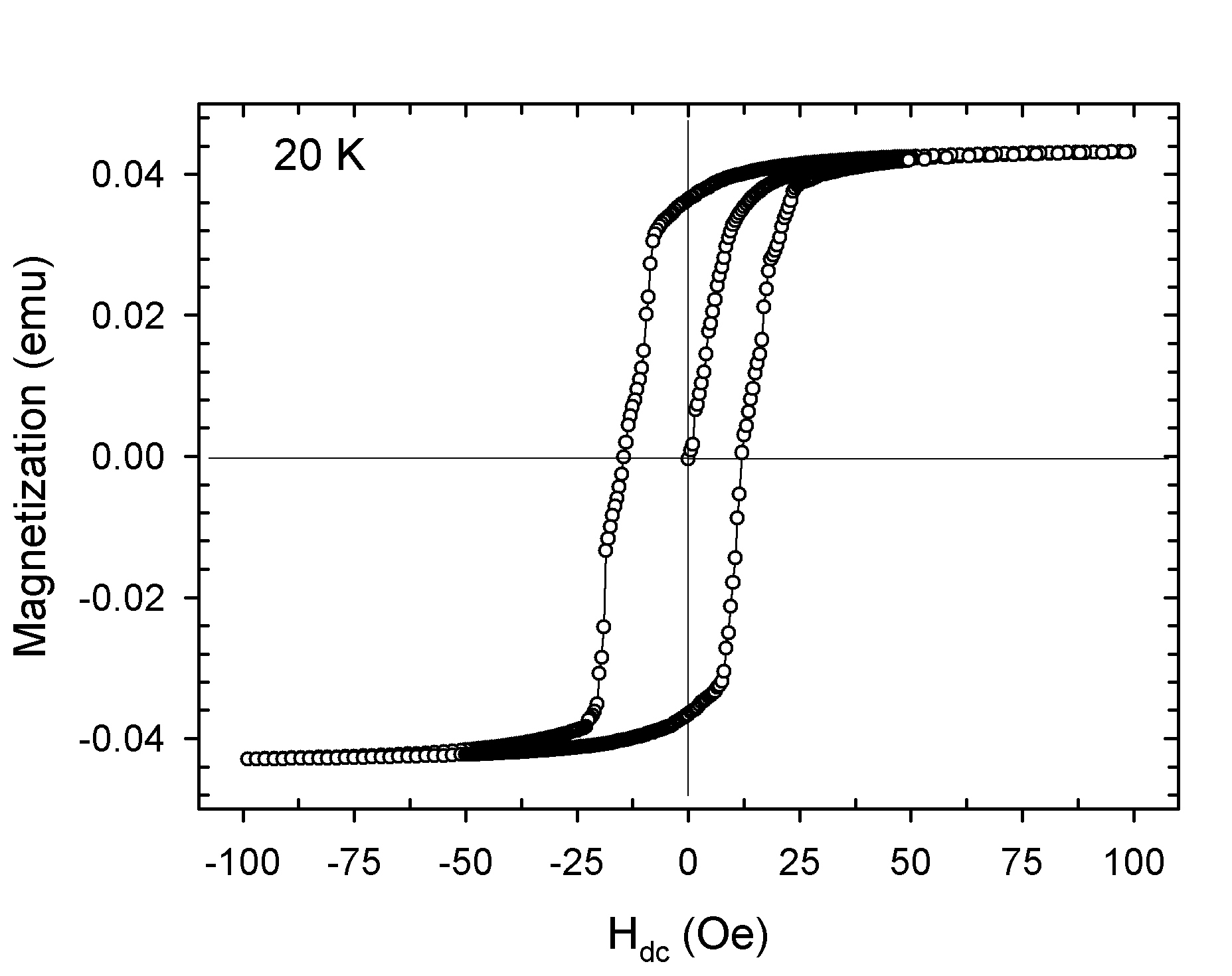} 
\caption{DC SQUID hysteresis of $\Co$ at 20 K. Field cycle time took approximately 20 hours.}
\label{squid1}
\end{figure}

In stronger fields, involved with major loops, there are different sources of the effective-$\tau$ reduction. First, the domain nucleation processes are now necessarily involved contributing with shorter relaxation time in the overall phase lag. Second, the frequency and field effects on the effective relaxation are actually mixed together: In the saturated (whisker-like) part of major loops there is no dynamics at all, as the domain walls had been ejected out of, now monodomain, sample. Dynamics exists only in the non-saturated parts of the loops which, at fixed field frequency, are ramped by the field with the sweeping rates proportional to the total field swing. As mentioned above, high sweeping rates (equivalently, high frequencies) select out the processes involving fast relaxations.

As far as the frequency dependence of the effective $\tau$ is concerned it can generally be interpreted along the same track- low frequency dynamics relies on processes with long relaxation times, typically involved with the pinned domain walls \cite{nat}. Important aspect of the frequency dependent $\tau_f(f)$ is that it naturally bridges our dynamic hysteresis loops with `static' hysteresis. Strictly speaking, dynamic $H_c$ vanishes in a true dc limit ($\omega \rightarrow 0$) in our model. However, any dc hysteresis can be more realistically and accurately represented as an ac one but taken in the low-frequency limit. For example, Fig.~\ref{squid1} shows the dc-SQUID  hysteresis data at 20 K. In taking this hysteresis loop the field cycle took approximately 20 hours, corresponding to frequency of $f={\rm 1.4\times  10^{-5}}$ Hz. The time spent on field cycling (at the order of hours) was found influencing the shape and the width (i.e., $H_c$) of `dc' hysteresis loops, as expected in models relying on thermal activation \cite{sha,sko1}. However, experimental value of $H_c^{dc}(\approx$ 20 Oe, Fig.\ref{squid1}) for mentioned $f$ is much bigger than the estimation for $H_c(\omega, \tau)$ on basis of Eq.(\ref{eq4}) for the corresponding choice of $\omega \tau$ (where $\omega=2 \pi f$ and $\tau$ are calculated as $\tau(f,H_0)=\tau_f(f) \tau_h(H_0)$ with the parameters specified in Fig.~\ref{fign4}). Obviously, the low-frequency behavior of $\tau_f(f)$ (for $f\ll 1$ Hz) cannot be easily extrapolated from studies of dynamic hysteresis in the frequency range $f\geq 1$ Hz). The reason is a complicated nonlinear physics of pinned domain walls \cite{nat,kle,che} which cannot be, at this stage, easily reduced to a one single phenomenological parameter, $\tau$. The triangle area in Fig.\ref{fign3} indicates a region in the $H_c(f)$ plot characterized by pronounced domain wall pinning effects. In the rest of the plot area the thermal-activation-enhanced Debye relaxation model, presented in this work, obviously perfectly interprets the dynamic hysteresis experimental data of $\Co$. 

In conclusion, we studied slowing down of the domain wall dynamics below the ferromagnetic transition of a new oxo-halide system $\Co$ by the induction-type dynamic hysteresis loops. Convenient quantitative matching of the dynamics' time scale with the available frequency window enabled detailed studies of minor loops in their dependences on frequency and applied field amplitude. The presented model for dynamic hysteresis loops relies just on the traditional ingredients of the spin-lattice relaxation theory and on the N\'eel-Brown thermal activation. The very low frequency deviations of experimental results from the model predictions are ascribed to specific physics of pinned domain walls, which remains outside scope of this work.

\section{Acknowledgments}

M.P., I. \v Z., D.D. and D.P. acknowledge financing from the projects 035-0352843-2845 and 119-1191458-1017 of the Croatian Ministry of Science, Education and Sport.


\begin{thebibliography}{99}

\bibitem{lee}T.H. de Leeuw, R. van den Doel, and U. Enz, Rep. Prog.Phys. {\bf 43}, 690 (1980).

\bibitem{nat}T. Nattermann,V. Pokrovsky, and V.M. Vinokur, Phys. Rev. Lett. {\bf 87}, 197005 (2001), and references therein.


\bibitem{kru}L. Krusin-Elbaum, T. Schibauchi, B. Argyle, L. Gignac, and D. Wellers, Nature {\bf 410}, 444 (2001), and references therein. 

\bibitem{sko1}see, e.g., R. Skomski, J. Phys.: Condens. Matter {\bf 15}, R841 (2003), and references threin.


\bibitem{fre}N.A. Frey and S.Sun in {\it Inorganic Nanoparticles: Synthesis, Applications, and Perspectives }, Editors C. Altavilla and E. Ciliberto, CRC Press (Taylor Francis Group), US (2010).


\bibitem{ses}R. Sessoli, Inorganica Chimica Acta {\bf 361}, 3356 (2008), and references therein. 

\bibitem{pan}Q.A. Pankhurst, J. Connolly, S.K. Jones, and J. Dobson, J.Phys.D: Appl.Phys. {\bf 36}, R167-R181 (2003).

\bibitem{fus4}With the focus on ferromagnetically ordered systems this short overview does not directly apply to equilibrium properties of molecular or single-chain magnets \cite{ses} (revealing no long range order thus no domain pattern). Dynamical aspects, presented throughout the article, are however equally relevant for the latter systems as well.

\bibitem{mor}See, e.g., A.H. Morrish, The Physical Principles of Magnetism, IEEE Press, (An IEEE Press Classical Reissue), New York, p.344 (2001).

\bibitem{sko}R. Skomski, J. Zhou, R. D. Kirby, and D. J. Sellmyer, J. Applied Phys. {\bf 99}, 08B906 (2006); R. Skomski, J. Applied Phys. {\bf 101}, 09B104 (2007).

\bibitem{moo}See., e.g., T. A. Moore and J. A. C. Bland, J. Phys.: Condens. Matter 16, R1369–R1386 (2004), and the references therein.

\bibitem{pop}I. S. Poperechny, Yu. L. Raikher, and V. I. Stepanov, Phys. Rev B 82, 174423 (2010)


\bibitem{pre}M. Prester, I. \v Zivkovi\'c, O. Zaharko, D. Paji\'c, P. Tregenna-Piggott, and H. Berger, Phys. Rev. B {\bf 79}, 144433 (2009).

\bibitem{bec}
R. Becker, M. Johnsson, H. Berger, M. Prester, I. Zivkovic, D. Drobac., M. Miljak, M. Herak,
Solid State Sci. {\bf 8}, 836 (2006).

\bibitem{kie}Specific heat measurements (K. Kiefer et al., unpublished) are fully consistent with the first order character of the ferromagnetic transition at $T_c$.

\bibitem{zen}
C. Zener, Phys. Rev. {\bf 96}, 1335 (1954).


\bibitem{col}K.S. Cole and R.H. Cole, J. Chem. Phys. {\bf 9}, 341 (1941).

\bibitem{fus1}Collapse of the different-temperature Cole-Cole plots is actually not expected as the isothermal susceptibility $\chi_T$ (see, text) should monotonically rise by lowering temperature \cite{che,kle}. We attribute the apparently temperature-independent $\chi_T$ to an artifact of uncertain demagnetizing field corrections. One namely notes a very high value of $\chi_T$ (close to 2000 emu/mol, corresponding to almost 60, in SI units), suggestive of demagnetizing factor corrections. However, the available data for the demagnetizing factor of homogeneously magnetized rectangular rods (in saturation) were found inapplicable for susceptibility corrections of our multi-domain sample (in low applied field). Thus the `intrinsic' Cole-Cole plot might differ to some extent from the experimental ones. In particular, the low-frequency asymmetry might be attributed to lack of the demagnetizing factor corrections.   


\bibitem{raq}B. Raquet, R. Mamy, and J.C. Ousset, Phys. Rev B {\bf 54}, 4128 (1996).


\bibitem{sha}M. P. Sharrock, J. Appl. Phys. {\bf 76}, 6413 (1994).


\bibitem{cher}See, e.g., N.A. Chernova, Y. Song, P.Y. Zavalij, and M.S. Whittingham, Phys.Rev. B {\bf 70}, 144405 (2004).


\bibitem{kra}S. Krause, G. Herzog, T. Stapelfeldt, L. Berbil-Bautista, M. Bode, E.Y Vedmedenko, and R. Wiesendanger, Phys. Rev. Lett. {\bf 103}, 127202 (2009).


\bibitem{koc}R.H. Koch, G.Grinstein, G.A. Keefe, Yu Lu, P.L. Troilloud, and W.J. Gallagher, Phys. Rev. Lett. {\bf 84}, 5419 (2000).


\bibitem{kle}W. Kleemann, J.Rhensius, O.Petracic, J.Ferr\'e, J.P.Jamet, and H.Bernas, Phys. Rev. Lett. {\bf 99}, 097203 (2007).

\bibitem{che}X. Chen, O. Sichelschmidt, W. Kleemann, O. Petracic, Ch. Binek, J. B. Sousa, S. Cardoso, and P. P. Freitas, Phys. Rev. Lett. {\bf 89}, 137203 (2002).


\bibitem{lab}M. Labrune, S. Andrieu, F. Rio, and P. Bernstein, J. Magn. Magn.
Mater. 80, 211 (1989).

\bibitem{kob}S. Kobayashi, S. Takahashi, T. Shishido, Y. Kamada, and
H. Kikuchi, J. Appl. Phys. 107, 023908 (2010)

\bibitem{fus}It is interesting to note that the same result for the relationship between the phase and the relaxation rate follows from the theoretically more rigorous approach \cite{leu} based on master equation for the kinetic Ising model within the scheme of Glauber dynamics \cite{gla}. 

\bibitem{leu}K. Leung and Z. Neda, Physics Letters A 246, 505-510 (1998).

\bibitem{gla}R. J. Glauber, J. Math Phys. 4, 294 (1963).


\bibitem{met}P.J.Metaxas, J.P.Jamet, A.Mougin, M.Cormier, J.Ferr\'e, V.Baltz, B.Rodmacq, B.Dieny, and R.L.Stamps, Phys. Rev. Lett. {\bf 99}, 217208 (2007). 




\end{thebibliography}
\end{document}